\documentclass
[twocolumn,amssymb]{revtex4}
\usepackage{epsfig}
\usepackage{hyperref}

\begin{document}
\title{The spinorial representation of polarized light and Berry phase}
\author{Dipti Banerjee}
\thanks{Regular associate of ICTP}
\email{deepbancu@homail.com,dbanerje@ictp.it}
\affiliation{Physics and Applied maths unit,\\Indian Statistical Institute\\ 203,B.T.Road,Kolkata-700035,\\ West Bengal, INDIA}

\begin{abstract}
From relativistic point of view it has been shown here that a polarized photon can be visualized to give an equivalent spinorial description when the two-component spinor is the eigenvector of $2\times2$ Hermitian, Polarization matrix. The Berry phase of the initial state can be calculated by matrix method as it complete one rotation over a closed path on the Poincare's sphere.
\end{abstract}
\maketitle

\section{Introduction}
It was about fifty years ago, the topological phase have been predicted theoretically first in optics. The parallel transport law for the polarization vector of an electromagnetic wave propagating through a medium with slowly varying refractive index has been studied by S.M.Rytov \cite{paper1} in 1938. He had shown that derivative of the rotation angle $\varphi$ with respective to the path length $s$ is equal to the torsion of the curved ray \cite{paper2}
\begin{equation}
 \frac{d\varphi}{ds}=\frac{1}{T}
\end{equation}
where $T$ is the torsion radius. This expression known as Rytov,s law. Later on,
in 1941 Vladimirskii \cite{paper3} predicted the topological phase of a curved non-planar light beam due to rotation of plane of polarization in an in homogeneous medium. The angle $\Theta$ due to this rotation is equal to the solid angle $\Omega$ subtended by the spherical contour $C$.
\begin{equation}
\Theta=\Omega(c)
\end{equation}
The angle coincides with the Berry phase for photons of fixed helicity found recently in the experiments of Tomita and Chiao \cite{paper4} in helically coiled fibre wave-guides.Chiao and Wu \cite{paper5} proposed to measure the Berry phase using effective optical activity of a coiled single-mode fibre waveguide.Berry in his paper \cite{paper6} predicted that any spin 1 particle, a boson can acquire a phase factor of $-1$ under certain rotations which classically return to its original states.

The earlier work by Pancharatnam \cite{paper7} carried out in 50,s shows similar phase for interference of polarized light represented by Poincare sphere.It was Ramseshan and Nityananda \cite{paper8} who pointed out the analogy between the phase of Pancharatnam and Berry. Later on, from quantum mechanical point of view Berry studied the topological phase for photon in linearly polarized light \cite{paper9} and helically coiled light \cite{paper10}. Berry discussed his adiabatic phase for polarized photon in comparison with Pancharatnam phase which is $-\Omega/2$ where $\Omega$ is the solid angle subtended by the area of the triangle on the Poincare sphere \cite{paper11} whose vertices are the three polarizations. Samuel and Bhandari \cite{paper12} reported first a direct experimental observation of the topological Pancharatnam-Berry's phase in the non-unitary evolution on the Poincare sphere by means of a laser interferometer.

It can be mentioned here that Pancharatnam phase is not identical with the optical phase given by the solid angle for coiled light propagating along a twisting path.Berry in his paper \cite{paper9} thus considered the photon in the plane polarized light traveling along the z direction as two component spinor,whereas in the coiled light experiement it acts as three component spinor.

Following the work of Berry we want to calculate relativistically the Berry phase for photon in a plane polarized light. As light passes through an anisotropic medium the quantum particle (photon) fixes its helicity.If positive helicity corresponds the right circular polarization then negative helicity belongs to the left circular polarization.
In analogy with a spin system, we may suggest that a photon with a fixed helicity can be viewed as if a "direction vector" $y_\mu$ is attached at the space-time point $x_\mu$ in the Minkowski space so that we can write the coordinate \cite{paper13} in a complexified space-time as $z_\mu=x_\mu+iy_\mu$. This leads us to treat a photon with a fixed helicity as a fermionic entity,the masslessness is ensured by the condition ${y_\mu}^2=0$.

In this paper, our aim is to study the Berry phase of a monochromatic polarized light from relativistic point of view over the triangle on the Poincare sphere whose vertices express three different nonorthogonal polarizations.
In section II, we shall relate the chiral photon as two-component massless spinor and in the section III our calculation of Berry's topological phase over a triangle ABC on Poincare sphere by matrix method will be shown.

\section{Polarized light and Spinorial Representation}

A monochromatic light is said to be polarized whenever it is transmitted through certain crystalline medium that allows electrical anisotropy. This indicates that photons in the
polarized beam fixes its helicity whose direction changes with the change of polarization axis. In an anisotropic space a particle having a fixed helicity can be viewed as a
direction vector attached at the space-time point \cite{paper13}.From relativistic point of view if $x_\mu$ be the position coordinate of the particle and $y_\mu$ indicates the direction vector then we can consider the position coordinate in the complex space as $z_\mu=x_\mu+iy_\mu$. This extended structure indicates the acquire of mass,the masslessness case is achieved at the condition $|y_\mu|^2=0$.

To relate this relativistic particle in the complex space as a spinor we can express its position coordinate in terms of spinorial variables in the matrix representation of the coordinate points.
Indeed, we can write the chiral coordinate as
\begin{equation}
z^\mu = x^\mu +iy^\mu
     = x^\mu +(i/2){\lambda_\alpha}^\mu  \theta_\alpha
\end{equation}
where $\alpha=1,2$
we have taken $y_\mu=i/2{\lambda_\alpha}^\mu \theta_\alpha$

If we now replace the chiral coordinates by their matrix
representation
\begin{equation}
z^{AA^\prime}=x^{AA^\prime}+(i/2){\lambda_\alpha}^{AA^\prime}\theta^\alpha
\end{equation}
where
\begin{equation}
x^{AA^\prime} = \left(
\begin{array}{rr}
x^0-x^1 & x^2+ix^3 \\ x^2-ix^3 & x^0+x^1
\end{array}
\right)
\end{equation}
with
$${\lambda_\alpha}^{AA^\prime} \in SL(2,c)$$
The introduction of additional degree of freedom $\theta,\bar{\theta}$
in the space-time geometry modifies the twistor
equation $(\bar{Z}_a Z^a=0)$ involving the {\it helicity} operator
\begin{equation}
S_{hel} = - {\lambda_\alpha}^{AA^\prime} {\theta}^\alpha {\bar
\pi_A}\pi_A^\prime
\end{equation}
which we identify as the internal helicity and corresponds to the
fermion number when the two opposite orientations of internal
helicities represent particle and antiparticle.
 It may be noticed that $\bar{\pi^A}\pi^{A^\prime}$ is the
spinorial variable corresponding the momentum $p^\mu$ conjugate to
$x_\mu$ through the relation
$p_\mu\longrightarrow p^{AA^\prime}=\bar{\pi^A}\pi^{A^\prime}$.
When we have ${p_\mu}^2 0$, the
particle will have its mass due to the non-vanishing character of the
quantity $|{y_\mu}|^2$. It is observed that the complex conjugate of
the chiral coordinate (3) will give rise to a massive particle with
opposite {\it internal helicity} corresponding to an antifermion. In
the null plane where $|{y_\mu}|^2=0$, we can write the chiral
coordinate for massless spinor as follows
\begin{equation}
z^{AA^\prime}=x^{AA^\prime}+\frac{i}{2}{\bar{\theta}}^A
\theta^{A^\prime}
\end{equation}
where the coordinate $y^\mu$ is replaced by
$y^{AA^\prime}=(1/2){\bar{\theta }}^A \theta^{A^\prime}$. In this
case the {\it helicity} operator is given by
\begin{equation}
S=-{\bar{\theta}}^A {\theta}^{A^\prime}{\bar \pi}_A
\pi_{A^\prime}=-{\bar{\varepsilon}}\varepsilon
\end{equation}
where $\varepsilon=i\theta^{A^\prime}\pi_A$,
$\bar{\varepsilon}=-i{\bar{\theta}}^A\pi_{A^\prime}$. The corresponding
twistor equation describes a massless spinor field. The state with
the internal helicity $+1/2$ is the vacuum state of the fermion
operator.
$$\varepsilon |S+1/2>=0$$.
 Similarly the state with internal
helicity $-1/2$ is the vacuum state of the fermion operator
$$\bar\varepsilon|S=-1/2>=0$$.

In case of massive spinor, we can define a plane $D^-$, where for
coordinate $z_\mu=x_\mu +iy_\mu$, $y_\mu$ belongs to the
interior of forward light-cone $(y>>0)$ and as such represents the
upper half-plane with the condition $dety^{AA^\prime}>0$ and
$(1/2)Try^{AA^\prime}>0$. The lower half plane $D^+$ is given by the
set of all coordinates $z_\mu$ with $y_\mu$ in the interior of the
backward lightcone $(y<<0)$. The map  $z \rightarrow z^*$ sends
the upper half-plane to the lower half-plane. The space M of null
planes $(det y^{AA^\prime}=0)$ is the Shilov boundary so that a
function holomorphic in $D^{-}(D^{+})$ is determined by its
boundary values. Thus,if we consider that any function
$\phi(z)=\phi(x)+i\phi(\xi)$ is holomorphic in the whole domain
the {\it helicity} $+\frac{1}{2}(-\frac{1}{2})$ in the null plane
may be taken to be the limiting value of the {\it internal
helicity} in the upper(lower) half-plane. Thus massless spinor
exists in this plane. For a massive particle {\it helicity} is
incorporated in the internal space and identified as {\it internal
helicity} which introduces solitonic feature for the spinor and
gives rise to a massive fermion. In view of this, the {\it
internal helicity} may be taken to represent the fermion number
where {\it helicity} is associated with spin for massless fermion.

We are motivated by the words of Berry \cite{paper17}, "Photons have no
magnetic moment and so cannot be turned with a magnetic field but
have the property of the helicity to use".This idea agrees with our
present explanation of photon with a fixed helicity in the polarized
light that can be viewed as a massless spinor with helicity $+1/2$
or $-1/2$ on the Shilov boundary. The eigen state of helicity operator with
eigenvalues $+1$ and $-1$ are referred to, respectively, as the {\it
right-handed} state (spin parallel to motion) and the {\it
left-handed} state (spin opposite to motion).


It may be noted that wave-function $\phi(z_\mu)=\phi(x_\mu)+i\phi(y_\mu)$
can be treated to describe a particle moving in the external space-time having the coordinate $x_\mu$ with attached{\it direction vector} $y_\mu$.Thus the wave
function should take into account the polar coordinates $r$,
$\theta, \phi$ along with the angle $\chi$ which specify the
rotational orientation around the {\it direction vector}
$y_\mu$. For an extended particle $\theta, \phi$ and $\chi$ just
represent the three Euler angles.

In a $3D$ anisotropic space,we can consider an axis symmetric
system where the anisotropy is introduced along a particular
direction. It is to be noted that in this anisotropic space, the
components of linear momentum satisfy a commutation relation of
the form
\begin{equation}
[p_i,p_j]= i\mu \epsilon_{ijk}\frac{x^k}{r^3}
\end{equation}

In such a axis-symmetric space,the conserved angular momentum $J$
is represented by
\begin{equation}
\vec{J}=\vec{r}\times \vec{p} - \mu\vec{r}
\end{equation}
which is similar to that of the angular momentum of a charged
particle moving in the field of a magnetic monopole. It follows
that $J^2=L^2-\mu^2$ instead of $L^2$ is a conserved quantity. In
general,$\mu$ which is the measure of the anisotropy given by the
eigenvalue of the operator $i\frac{\delta}{\delta\chi}$ and can
take the values $\mu=0,\pm 1/2,\pm 1,\ldots$.The spherical
harmonics incorporating the term $\mu$ be written as \cite{paper15}

$${Y_l}^{m,\mu}=(1+x)^{-(m-\mu)/2}\,(1-x)^{-(m+\mu)/2}$$
\begin{equation}
\frac{d^{l-m}}{{dx}^{l-m}}\,[(1+x)^{l-\mu}(1-x)^{l+\mu}\,]\times
e^{im\phi}e^{-i\mu\chi}
\end{equation}
with $x=cos\theta$.

In the anisotropic space a scalar particle moving with $l=1/2$
with $l_z=+1/2$ can be treated as a spinor with {\it helicity}
$+1/2$.The specification of the $l_z$ value for the particle and
antiparticle states then depicts it as a chiral spinor. This gives
for $m=\pm 1/2,\mu=\pm 1/2,$ the following spherical harmonics from
the relation (11) in terms of the components $(\theta,\phi,\chi)$

\begin{equation}
\left.
\begin{array}{lcl}
{Y_{1/2}}^{1/2,1/2} & = & \sin \frac{\theta}{2} e^{i(\phi-\chi)/2} \\
{Y_{1/2}}^{-1/2,1/2} & = & \cos \frac{\theta}{2} e^{-i(\phi+\chi)/2} \\
{Y_{1/2}}^{1/2,-1/2} & = & \cos \frac{\theta}{2} e^{i(\phi+\chi)/2} \\
{Y_{1/2}}^{-1/2,-1/2} & = & \sin \frac{\theta}{2}
e^{-i(\phi-\chi)/2}
\end{array}
\right\}
\end{equation}

These represent spherical harmonics for half-orbital angular
momentum in an anisotropic space specified by commutation relation of the linear momentum components given by equation (10) with $\mu=\pm 1/2$.It is noted from these spherical harmonics, we can construct the product wave functions which can be formulated in the standard coordinate coordinate representation where anisotropic character suppressed due to cancelation of $\mu$ in the harmonics. In fact we have the following spherical harmonics

${Y_{1/2}}^{1/2,1/2}.{Y_{1/2}}^{1/2,-1/2}={\sin\theta/2}{\cos\theta/2}e^{i\phi}$\\
$=\sin\theta e^{i\phi}={Y_1}^1.$\\ \\\emph{}
${Y_{1/2}}^{-1/2,1/2}.{Y_{1/2}}^{-1/2,-1/2}={\cos\theta/2}{\sin\theta/2}e^{-i\phi}$\\
$={\sin\theta}e^{-i\phi}={Y_1}^{-1}.$\\
\begin{equation}
{Y_{1/2}}^{-1/2,1/2}.{Y_{1/2}}^{1/2,-1/2}-{Y_{1/2}}^{1/2,1/2}.{Y_{1/2}}^{-1/2,-1/2}
\end{equation}
$=\cos\theta ={Y_1}^0.$\\
in the product space in which the sphere is generated by $r,\theta$ and $\phi$ parameters.
In the next section we shall express the polarization matrix of the optical device in terms of the spherical harmonics in the product space. Afterwards the two component spinorial form of the chiral photon is chosen properly in terms of spherical harmonics
$\theta,\phi$ and $\chi$.

\section{Polarized Light and Berry Phase}
The propagation of light through an optical system was studied by Jones\cite{paper17}
in a $2X2$ matrix method. The method was based on the idea that in anisotropic media, the displacement vector D that represent the elliptic vibration or light can be represented by the column vector
\begin{equation}
\vec{D}={D_1 \choose D_2}
\end{equation}

where $D_1$ and $D_2$ are the resolved components of the electric displacement vector D along OX and OY are in general complex numbers. The intensity is $I=|D_1|^2 + |D_2|^2 $
while the complex ratio $D_2/D_1$ describes its polarization state.

The passage of light through an optical component such as birefringent,absorbing or dichroic plate would be to change both $D_1$ and $D_2$ so that the effect may be represented by a $2\times2$ matrix with complex elements \cite{paper18}
\begin{equation}
\vec{D}'=M\vec{D}
\end{equation}
For a non-absorbing plate,there is no change in the intensity and the matrix M is therefore unitary i,e. $detM=1$ which makes $|D'|=|D|$.

On similar manner Berry \cite{paper9} pointed out for a monochromatic wave               traveling in the z direction, the polarization state of the electric displacement vector lying in the XY plane becomes
\begin{equation}
|\psi>={{\psi_+} \choose {\psi_-}}
\end{equation}
where $\psi_{\pm}=(d_x \pm id_y)/\sqrt{2}$ a two component spinor.Each such $|\psi>$ is the eigenvector with eigenvalue $+1/2$ of the polarization matrix of the form
\begin{equation}
\left.
\begin{array}{lcl}
H(r)&=&{\vec{r}}.{\vec{\sigma}}\\
&=& \frac{1}{2}\pmatrix{z &(x-iy) \cr (x+iy) & {-z}}\\
&=& \frac{1}{2}\pmatrix{\cos\theta & \sin\theta e^{-i\phi}\cr
{\sin\theta e^{i\phi}}&{-\cos\theta}}
\end{array}
\right\}
\end{equation}

where $\sigma$ is the vector of Pauli spin matrices and $r(x,y,z)$ is a unit vector with angles $\theta$ and $\phi$. The sphere with coordinates $r,\theta,\phi$ is the Poincare sphere whose each point define a matrix H. All the matrix on the equatorial circle of sphere represent linear polarizations. The matrix at the two poles (North and South) represent the left and right circular polarization respectively.Besides these all other points on the surface of Poincare sphere represent the elliptic polarization with respective sense of orientation in the upper or lower domain.

The polarization matrix $H(r)$ can be constructed from the two component spinor $|\psi>$ through the relation $(|\psi><\psi|-1/2)$.

It implies that the proper eigenvector may be chosen from the Hamiltonian matrix as follows when we choose $|\psi>=\frac{1}{\sqrt{2}}{\psi_+ \choose \psi_-}$ with $\psi_\pm=d_x \pm id_y$.
\begin{equation}
\left.
\begin{array}{lcl}
H(r)&=& \frac{1}{2}{\psi_+ \choose \psi_-}(\psi_+~~~\psi_) - \frac{1}{2}\\
&=& \frac{1}{2}\pmatrix{{\psi_+\psi_+} & {\psi_+\psi_-} \cr {\psi_-\psi_+} & {\psi_-\psi_-}} -\frac{1}{2}\pmatrix{1 & 0 \cr 0 & 1}\\
&=& \frac{1}{2}\pmatrix{{\psi_+\psi_+ -1} & {\psi_+\psi_-} \cr {\psi_-\psi_+} & {\psi_-\psi_- -1}}
\end{array}
\right\}
\end{equation}
Each element in this $2X2$ matrix can be replaced by product wave functions as shown in the previous sections
\begin{equation}
\left.
\begin{array}{lcl}
H(r)&=& \frac{1}{2}\pmatrix{\cos\theta & {\sin\theta e^{-i\phi}} \cr {\sin\theta e^{i\phi}}& -\cos\theta}\\
&=& \frac{1}{2}\pmatrix{{Y_1}^0 & {Y_1}^1 \cr {Y_1}^-1 &{Y_1}^0\cr}\\
&=&\frac{1}{2}\pmatrix{({Y_{1/2}}^{-1/2,1/2}.{Y_{1/2}}^{1/2,-1/2}-{Y_{1/2}}^{1/2,1/2}.{Y_{1/2}}^{-1/2,-1/2})
&{Y_{1/2}}^{-1/2,1/2}.{Y_{1/2}}^{-1/2,-1/2}
\cr {Y_{1/2}}^{1/2,1/2}.{Y_{1/2}}^{1/2,-1/2} &({Y_{1/2}}^{1/2,1/2}.{Y_{1/2}}^{-1/2,1/2}-{Y_{1/2}}^{1/2,1/2}.{Y_{1/2}}^{-1/2,-1/2})}\\
\end{array}
\right\}
\end{equation}
where $${Y_{1/2}}^{1/2,1/2}.{Y_{1/2}}^{-1/2,-1/2} \approx 1$$
on comparing the equations (18) and (19) we have
$${Y_1}^1 \approx \psi_- \psi_+ \approx {Y_{1/2}}^{-1/2,1/2}.{Y_{1/2}}^{-1/2,-1/2}$$
$${Y_1}^{-1} \approx \psi_+ \psi_- \approx {Y_{1/2}}^{1/2,1/2}.{Y_{1/2}}^{1/2,-1/2}$$
and
$${Y_1}^0 \approx (\psi_+ \psi_+ -1) \\or (1-\psi_- \psi_-) \approx\\ {Y_{1/2}}^{1/2,1/2}.{Y_{1/2}}^{-1/2,-1/2}\\-{Y_{1/2}}^{-1/2,1/2}.{Y_{1/2}}^{1/2,-1/2}$$
Lets us choose among these only one (say former) pair as the eigenvector in terms of spherical harmonics to represent a relativistic spinor.
Thus
\begin{equation}
|\psi>={\psi_+ \choose \psi_-}={{Y_{1/2}}^{1/2,1/2} \choose {Y_{1/2}}^{1/2,-1/2}}
= {{\sin\theta/2 e^{i(\phi-\chi)/2}} \choose {\cos\theta/2 e^{i(\phi+\chi)/2}}}
\end{equation}
It can be noticed here that the Poincare sphere which represents polarization matrices lie on the product space of the space spanned by the eigenvectors.

This space of eigenvectors is extended in nature due to the introduction of anisotropy by the additional degree of freedom $\chi$.

In the work of Berry \cite{paper9} the polarization state of displacement vectors are defined on complex plane. We consider two dimensional plane as the surface of three dimensional sphere and thus an extra parameter is introduced in the spinor structure.

The equation (20) which represent a chiral photon as two-component relativistic spinor can be splitted as follows.
\begin{equation}
{\psi_+ \choose \psi_-}={\psi_+ \choose 0}+{0 \choose \psi_-}
\end{equation}
where the first term in the right hand side represent the eigenvector in the upper hemisphere and the second term denotes the vector of lower hemisphere for left and right polarization respectively. For linear polarization the eigenvector consists both $\psi_+$ and $\psi_-$ but in addition here the angle $\chi$ which helicity makes with the complex plane becomes zero. With this idea \cite{paper9}, at the poles of extended sphere $\theta=0$ and $\theta=\pi$ which gives $\psi_+=0$ and $\psi_-=0$ representing two circular polarization respectively.For the points on the equitorial circle $\theta=\pi/2$ indicates $|\psi_+|=|\psi_-|$ represent the different linear polarization where orthogonal points are situated at $\phi=\pi$ apart.

\vspace{10cm}

It has been pointed out recently by Ramaseshan and Nityananda and afterwards Berry that there is an intimate analogy between the Berry phase and the Pancharatnam phase which is developed as the polarization state of the light beam completes a closed circuit on the Poincare sphere.The difference lies in their approach, since Berry to find his topological phase considered the polarized light as two-state quantum system(a spin -1/2 particle) whose evolution is governed by the Hamiltonian operator in equation (18).According to Pancharatnam if a polarized beam is divided into two
components in which one continues in the state $|A>$ in state $|A'>$ via intermediate states $|B>$ and $|C>$,then interference between $|A>$ and $|A'>$ express the phase difference $-\Omega_{ABC}/2$.The relation being
\begin{equation}
<A|A'>= exp(-i\Omega_{ABC}/2)
\end{equation}
where $\Omega_{ABC}$ is the solid angle of the triangle ABC.From the
quantum mechanical point of view,at the end of the closed circuit traversed by a
two-component spinor which is here identified as polarized light the Berry phase
becomes
\begin{equation}
<A|A'>=exp^{-i\Omega(C)/2}
\end{equation}

In a relativistic framework,we want to calculate the Pancharatnam-Berry phase
of a linearly polarized monochromatic light traveling along z direction and
completing a closed circuit ABCA on the Poincare sphere.The points A and C situated on the equitorial circle represent two linear polarizations at two different value of $\phi$ say $\phi=0$ and $\phi=\pi/2$ respectively. The z axis is assumed passing
through the point B at the pole that representing circular polarization. Considering
any one of the three points we can calculate by Jones matrix \cite{paper17} method
the topological phase of polarized photon during one complete rotation along the circuit. It must be noted \cite{paer12} that for the sphere of directions,the topological phase is given by $\Omega$, whereas for the Poincare sphere, it is given by $\Omega/2$ as in the case of Berry,s phase for spin $-1/2$ rotations.

Let us choose the initial point be the pole B then the initial eigenvector given by coordinate $\theta=0$ and $\phi=\pi/2$
\begin{equation}
|\psi_B>={0 \choose e^{i(\chi+\pi/2)/2}}
\end{equation}
which represent only the helicity $+1/2$ for say left circular polarization. The
polarization matrix equation (15) at the point B for $\theta=0$ and $\phi=\pi/2$.
\begin{equation}
H_B=\pmatrix{1 & 0\cr 0 & {-1}}
\end{equation}
On similar manner the polarization matrix for points a and c at the respective coordinates $\theta=\pi/2,\phi=0$ and $\theta=\pi/2,\phi=\pi/2$ becomes
\begin{equation}
H_A=\pmatrix{0 & 1\cr 1 & 0},H_C=\pmatrix{0 & -i\cr i & 0}
\end{equation}
The final eigenvector $|\psi_B'>$ at the point B is obtained as the chiral photon completes the rotation along the path BCAB. Identifying the vector $|\psi_B>$ as Jones vector of the incident beam then $|\psi_B'>$ can be calculated on multiplication with
$|\psi_B>$ the matrices. $H_C,H_A$ and $H_B$ which represent three homogenous optical devices \cite{paper19}.
$$|\psi_B'>=H_B H_A H_C|\psi_B> $$
$$|\psi_B'>=\pmatrix{1&0 \cr 0&{-1}}\pmatrix{0&1 \cr 1&0}\pmatrix{0&{-i} \cr i&0}
{0 \choose e^{i(\chi+\pi/2)/2}}$$
$$=\pmatrix{0&1 \cr {-1}&0}\pmatrix{0&{-i} \cr i&0}{0 \choose e^{i(\chi+\pi/2)/2}}$$
$$=\pmatrix{i&0 \cr 0&i}{0 \choose e^{i(\chi+\pi/2)/2}}$$
\begin{equation}
|\psi_B'>={0 \choose ie^{i(\chi+\pi/2)/2}}
={0 \choose e^{i(\chi+3\pi/2)/2}}
\end{equation}

The scalar product between $|\psi_B'>$ and $|\psi_B>$ is our required geometrical phase which the two state quantum photon acquire over the path BCAB
\begin{equation}
<\psi_B|\psi_B'>=[0 e^{-i(\chi+\pi/2)}]{0 \choose e^{i(\chi+3\pi/2)/2}}=e^{i\pi/2}
\end{equation}
It seems from our result that for solid angle $\Omega=-\pi$ subtended by geodesic triangle ABC at the origin O,the Berry phase for helical photons is $\pi/2$. This gives
the explanation that a complete rotation when subtends the solid angle $\Omega-2\pi$,
the field of photon reverses.For polarized light the solid angle is the analogue of
magnetic flux to an abstract monopole \cite{paper9} of strength $-1/2$ at the center of the sphere.

\section{Discussion}
We have shown above that a polarized photon can be treated a two component spinor in a spinorial representation which helps us to relate Berry phase with the polarization phenomenon.In a complexified space-time having an anisotropy, a spinor can be viewed as a scalar particle moving with half orbital angular momentum having a fixed $L_z$ value $+1/2$ and $-1/2$ that can be related with the helicity of photon. The rotation of two state quantum photon over a closed path even in the absence of adiabaticity and unitary gives rise to the topological phase which is the celebrated Berry phase. This phase reveals the solid angle of the area enclosed on the Poincare sphere. Berry has pointed out the analogy of this phase with the magnetic flux such that a fictitious monopole of strength $-1/2$ is situated at the origin. This implies that an interaction of polarized photon or equivalently a spin $1/2$ particle with the fictitious monopole is assumed.We may note here that the behavior of angular momentum of a particle in an anisotropic space will be similar to that of a charged particle moving in the field of a magnetic monopole \cite{paper13}.

Finally we may point out that in a recent paper \cite{paper20}, we have shown the relationship of the berry phase with the chiral anomaly and the topological origin of
fermion number. The spinorial formulation of polarized light suggests that in the case of a fermion as well as in polarized light,the Berry phase has its common origin.

\end{document}